# Prediction and accelerated laboratory discovery of previously unknown 18-electron ABX compounds

Romain Gautier[1,†], Xiuwen Zhang[2,†], Linhua Hu[1], Liping Yu[2], Yuyuan Lin[1], Tor O. L. Sunde[1], Danbee Chon[1], Kenneth R. Poeppelmeier[1,*], Alex Zunger[2,*]

[1]Department of Chemistry, Northwestern University, 2145 Sheridan Road, Evanston, Illinois 60628-3113, United States

[2]University of Colorado, Boulder, Colorado 80309, United States

Communicating Author: K. R. Poeppelmeier and A. Zunger

[†] R.G. and X.Z. are equally contributing authors

Chemists and material scientists have often focused on the properties of previously reported compounds, leaving out numerous unreported but chemically plausible compounds that could have interesting properties. For example, the 18-valence electron ABX family of compounds includes the half-Heusler subgroup, and features examples of topological insulators, thermoelectrics and superconductors, but only 83 out of 483 of these possible compounds have been made. Using first-principles thermodynamics we have examined the theoretical stability of 400 unreported members and predict that 54 should be stable. 15 previously missing materials, now predicted stable, were grown in this study; X-ray studies agreed with the predicted crystal structure in all 15 cases. Among the characterized properties of the missing compounds are potential transparent conductors, thermoelectric materials and topological semimetals. This integrated process-prediction of functionality in unreported compounds followed by laboratory synthesis and characterization-could be a route to the systematic discovery of hitherto missing, realizable functional materials.

*The problem of missing Compounds*: An interesting observation surrounding current compilations of structures of previously synthesized inorganic compounds is the relatively large proportion of chemically reasonable atom combinations that are *not reported* therein[1-3]. Some of these "missing compounds" can be readily rationalized as being unstable for chemically obvious reasons, but it is difficult to speculate why many of the others are unreported. Not only could the structure and properties of such missing compounds (should they exist) improve our understanding of chemical trends within series of materials currently featuring missing entries, but the history of solid state chemistry and material science suggests that new compounds come with new and potentially technologically relevant functionalities. Laboratory discovery of new functional materials without prior information to narrow down the likely compositions and structures can be challenging: In addition to the large number of possible combinations of elements, unknown crystal structures can complicate the process of identification and structure determination, as one cannot *a priori* classify and identify a set of non indexed diffraction peaks as belonging to a specific new phase. As a consequence, many synthetic attempts, so called *"dark reactions"* which are never reported[4], are carried out to identify a new phase among known and unknown phases and to isolate it for further structural characterization.

*Different theoretical approaches to discovery of functional compounds:* The widespread accessibility of first-principles electronic structure codes and fast, multi-processor computational platforms has proliferated the prediction of previously unknown thermodynamically stable compounds[5-8] as well as separate predictions of functionalities of materials. Three main styles of literature predictions are noteworthy: (i) The prediction of *unknown properties[9-16] of generally known, previously synthesized[1,2] compounds.* This includes high throughput calculations of properties of materials listed in databases[10,11,14-17]. Here stability is generally taken for granted, given that the material and its crystal structures are known experimentally. (ii) The prediction of *unknown properties of artificial, non-equilibrium structures*[18-22] (such as superlattices; core-shell nanostructures.) Here the issue of thermodynamic stability does not arise because activation barriers are presumed to be generally insurmountable under ordinary conditions. (iii) The forecast of new properties of hypothetical compounds in presumed structures[23-27]. Here, authors generally do not examine the theoretical stability the crystal structure as their interest lies in the properties of a postulated structure. Approach (i) runs the risk of missing potentially stable but previously overlooked unreported compound, and approach (ii) relies on protection by kinetic barriers. Approach (iii) can deliver exciting predicted properties albeit in potentially unstable structures, unrealizable in practice (e.g., high bulk modulus in PtN[26]; optoelectronics[24], hypothetical topological insulation[27], and piezoelectricity[25] in some of the hypothetical half-Heusler compounds). Also, "model Hamiltonian" approaches that specify the Hamiltonian via generic interaction terms (such as spin-spin or electron-electron or electron-phonon) but do not specify the chemical and structural identity of the system (atomic numbers, composition and structure) might predict interesting generic properties but lack a bridge to the materials in which such predicted properties will "live".

Here we apply another theoretical approach followed by laboratory examination of missing compounds. Regarding the *predictive theory*, we examine groups of unreported, ("missing") compounds via first principles thermodynamics, an approach in which the Hamiltonian is given atomic numbers, composition and structure and seeks compounds that are stable statically and dynamically with respect to competing phases. The calculation separates the 'unreported-predicted-stable' from the 'unreported-predicted-unstable' compounds, and then forecast material properties of the stable new materials in their affirmed structures. Validation of the method is conducted by applying it to previously reported compounds from the same general chemical class, and verifying that in all cases stability and the observed low T structures are correctly predicted.

This approach combines true discovery of previously missed compounds with prediction of properties that are expected to live in chemically recognizable, stable structures. Unlike approach (i) we focus on hitherto missed compounds, and unlike approaches (ii) and (iii) we filter out manifestly unstable compounds before advertising their interesting new properties. Unlike Model Hamiltonian approaches we chemically and structurally identify the compound in which the predicted functionality lives. Regarding the *laboratory examination*, we build on the aforementioned theory that (a) determines which compounds are stable, (b) predicts their crystal structures, thereby providing information that allows chemists to perform directed rational synthesis and structure identification in multiphasic samples as well as structure refinement; and (c) allows one to select the candidate compounds exhibiting the most interesting properties. We will illustrate how the iteration between experiments and theory greatly accelerates the discovery of novel materials with interesting functionalities.

*An interesting family of compounds:* We illustrate this approach to discovery of missing functional materials by considering the 18-valence electron ABX system. We focus here on the 1:1:1 ABX family where many interesting properties have been identified[17,24,25,27-32]. This family includes the "half-Heusler" compounds[33] as well as numerous compounds that are not cubic. These compounds are a conceptual extension of the well-known $s^2p^6$ (octet) binary semiconductors, such as I-VII (e.g., NaCl), II-VI (e.g., ZnSe), III-V (e.g., GaAs), or IV-IV (e.g., SiC), obtained by insertion of a group-X element (Ni, Pd, Pt) into an 8-valence electron lattice (Fig. 1), or by inserting a group-IX element (Co, Rh, Ir) into the 9-valence electron lattice (Fig. 2). Curiously, however, of the 483 ABX 18-electron materials that could be conceived from these groups, only 83 were previously made (check marks in Figs. 1 and 2); the remaining 400 are unreported in compilations of synthesized systems, such as Refs. [1-3] as well as in additional journal publications known to us[34-36]. In fact, the entire I-X-VII, II-X-VI, II-IX-VII, and III-IX-VI sub groups of the 18-electron ABX family are not documented. Yet, the known compounds from these ABX groups include interesting functionalities such as thermoelectricity[31], superconductivity[37], piezoelectricity[25], topological band structure properties (mostly from the III-X-V group)[27-30], and *p*-type transparent conductivity[38], raising the prospects that the missing compounds from these groups may have interesting properties.

*Determination of stability vs instability of hitherto missing compounds:* Using first principles density functional theory, we examine the thermodynamic stability of each missing 18-electron ABX compound with respect to (i) other crystal structures (including different stoichiometry), (ii) stability with respect to decomposition into any combination of their constituents and (iii) dynamic (phonon) stability. We address stability of a missing ABX compound under ambient pressure within density functional theory (DFT) in two steps (1) determine the lowest-energy crystal structure among many possibilities which is a dedicated problem by itself, both via genetic algorithm and via a fixed list of structure types (Supplementary Tables S1-S2), including phonon stability and (2) compare the total energy of the lowest-energy ternary structure to all the combinations of competing phases and to other ternary compounds in few other compositions; The protocol of testing the stability of a hypothetical compound described here applies to the ground state (T=0) structure, sorting out the stable ground states from those that are not ground states because specific competing phases or dynamical instability precludes their existence. If a compound is proved to be a ground state on the Convex Hull, it must appear in the composition-temperature phase diagram. Naturally, the temperature in which it will appear may or may not be experimentally accessible, and a complete, finite T calculation of phase diagram could be desirable (e.g., those predicted in Refs.[39-41]). We did not attempt phase diagram calculations here, an effort that might be done in the future. Interestingly, when applying our T=0 protocol to 44 compounds reported in the ICSD ($Mn_2SiO_4$, $Sr_2TiO_4$, $Al_2ZnS_4$, $Ba_2TiS_4$, $Ca_2SiS_4$, $Sc_2MgSe_4$, $In_2MgTe_4$ [42], BaZnSi, BaZnSn, CaZnGe, AgKO, KCaBi, CuKSe, KMgAs, KMgP, KZnSb, LiAlGe, LiAlSi, LiBeN, CuLiO, LiInGe, LiMgN, LiSrSb, LiYGe, NaAlSi, RbCaAs, SrZnSi [43], TiPtGe, VCoSi, VCoGe, NbCoSi, NbCoGe, NbCoSn, NbRhSi, NbRhGe, NbRhSn, NbIrSi, NbIrGe, NbIrSn, TaCoGe, TaRhGe, TaCoSi, TaRhSi, and TaIrSi), we find in all cases that we correctly predict their stability and in the correct (observed) structure. This suggests that the T=0 protocol is at least sometimes if not often a good predictor of laboratory existence in the said structure, from these closed-shell inorganic groups, and that many such previously made compounds reported in the ICSD represent a stable ground state, not a metastable structure. We note, in passing, the importance of examining groups of candidate competing structure, as, for example, the study of Ref. [44] has overlooked a large number of unreported ABX compounds as well as previously synthesized materials. These structures were disqualified from being stable because only a cubic form (often not the ground state) was allowed, leading to the very low yield (< 0.1%) of material prediction out of the initial set examined in addition to the assignment of instability on a large number of previously synthesized materials.

*Previously missing now predicted stable compounds:* Of the 400 missing compounds, we predict 54 to be stable in specific structures ('+' symbols in Figs. 1 and 2) and the remaining 346 to be unstable in all structures examined ('-' symbols in Figs. 1 and 2). The stability of four compounds are too close to determine as judged by the energy distance to the convex hull (see Fig. S1c). The predicted stable 18-electron ABX compounds, their lowest-energy crystal structure (see more details in Supplementary Tables S3-S8), and formation enthalpy are given in Table 1. The closest competing phases are given in the Tables S9 and S10 in Supplementary Information.

*Predicted compounds in new groups vs predicted compounds that supplement previously known groups*: It is interesting to see whether certain nominal chemical groups are missing for a good reason. Groups I-X-VII, II-X-VI, II-IX-VII, III-IX-VI, and subgroup III-X-V (III = Al, Ga, In) are thus included in our study even though they previously did not contain a single reported member, i.e. being missing groups. We predict 10 stable compounds from subgroup III-X-V (III = Al, Ga, In) (GaNiBi, InNiSb, InPdSb, InPtSb, GaNiSb, GaPtSb, AlNiSb, AlPtSb, AlNiAs, and AlNiP), 3 from group III-IX-VI (GaIrTe, AlIrSe, and ScRhTe), and 1 from II-X-VI (MgPdTe). The remaining 40 previously missing now predicted stable compounds belong to groups containing some known members. For example, 12 out of the 27 compounds in the (Ti, Zr, Hf)(Co, Rh, Ir)(P, As, Sb) group were missing, and 11 of them are now predicted to be stable (TiRhP, TiIrP, HfRhP, HfIrP, TiIrAs, ZrIrAs, HfRhAs, HfIrAs, TiIrSb, ZrIrSb, and HfIrSb), making the sub-group almost complete. The exception is the compound ZrCoAs which is slightly unstable with an energy distance to the strongest competing phases (or convex hull) $\Delta H_f(ABX)$-$C(x_{ABX})$ (see Methods section) of only 11 meV/atom. Also, 5 out of 18 compounds in the (V, Nb, Ta)(Co, Rh, Ir)(Si, Ge) group were missing, and all of them are now predicted to be stable (VRhSi, VIrSi, VRhGe, VIrGe, TaIrGe), thus completing the group.

*Predicted trends in crystal structure types*: The crystal structures of the predicted ABX compounds are given in Table 1. The labels of structure types are taken from Ref.

[43] and are listed for convenience in Supplementary Tables S1-S2. We denote in Figs. 1 and 2 compounds with the cubic half-Heusler structure (LiAlSi-type, F-43m) by violet bold symbols. We note some *emerging structural trends:* It is now clear that ABX compounds with light atom X (i.e. O, S, Se, N, P, As, C, Si, and Ge) tend to have non-cubic structures, while ABX compounds with heavy X (i.e. Te, Sb, Bi, Sn, and Pb) tend to have the cubic (also called half-Heusler) structure. For example, the predicted ABX compounds in groups IV-X-IV, IV-IX-V, and V-IX-IV with heavy X elements Sb, Bi, Sn, or Pb are all cubic.

*Previously missing now predicted unstable compounds*: We predict that as many as 346 (85%) (4 are borderline cases) of the 400 missing ABX compounds are thermodynamically unstable (shown by minus signs in Figs. 1 and 2), so they are "missing" for a good reason. Most of them are *unstable because of competing multinary compounds* (e.g. CaNiO is unstable w.r.t. CaO+Ni). A few are *unstable with respect to decomposition into elemental phases* (i.e. $\Delta H_f > 0$): AuPdCl ($\Delta H_f$ = 0.03 eV/atom), AuPtCl (0.12 eV/atom), InNiN (0.08 eV/atom), InPdN (0.14 eV/atom), InPtN (0.15 eV/atom), VCoPb (0.17 eV/atom), VRhPb (0.03 eV/atom) and a borderline case VRhC (0.01 eV/atom). Some of these compounds, now predicted to be unstable, (e.g. CaNiO, SrPtSe, and BaPdTe) were the subject of a previous theoretical study predicting interesting physical properties[25]. These are unlikely to materialize in standard growth methods.

*Trends in thermodynamic stability when varying anions or cations:* We find that all carbides and almost all nitrides (except LaNiN which is slightly stable) are thermodynamically unstable, in sharp comparison with the number of stable compounds with heavier anions that are less ionic, indicating that ionicity is not preferred in these 18-electron ABX compounds. Comparing the two sub-groups of III-X-V with III = Al, Ga, In vs III = Sc, Y, La, we find that the former subgroup was all missing and after our prediction still has fewer stable compounds than the latter subgroup. The key factor is that in the III-X-V group with III = Sc, Y, La there are two transition metal species whose *d* orbitals hybridize and repel, displacing the occupied *d* bands to lower energy thus stabilizing the compound. In contrast, the former subgroup contains only one transition metal specie and therefore lacks the energy stability from *d-d* bonding.

*Laboratory synthesis of never before made compounds now predicted stable:* Our first synthetic targets in the challenge of experimental verification were those compounds in the two most populated structure types, the cubic LiAlSi-type and orthorhombic MgSrSi-type structures (see Supplementary Table S1). Among these groups, many compounds were also not attempted owing to reagent toxicity (e.g. arsenides) or weak predicted thermodynamic stability (i.e. energetic distance from the convex hull). Attempts to synthesize the predicted stable compounds were made using arc melting and quartz tube annealing, two of many synthetic techniques available to make new materials. Figure 3 shows the realized single phasic *vs* multi phasic compounds. The compounds HfIrSb, TaIrSn, ZrIrSb, TiIrSb, ZrNiPb and HfRhP were made *single phase* and the X-ray diffraction (XRD) pattern (see Figs. 4 and S6-S8) clearly validates the predicted structure. It is interesting to note that single phases made so far have $\Delta H_f(ABX)-C(x_{ABX}) < -0.13$ eV/atom while VIrSi and ZrRhBi with $\Delta H_f(ABX)-C(x_{ABX}) \leq -0.13$ eV/atom were made with high purity in a *multi phasic* products. The other made *multi phasic* compounds ScRhTe, ScPtBi, VRhSi, ScPdP, ZrIrBi, TiIrP and ZrPdPb have weaker thermodynamic stability and were selected to test the theoretical predictions. We learned that certain compounds with very weak stability are hard to make even in multi phasic mixtures by the fact that all attempts to synthesize AlNiP and InNiBi were unsuccessful. Transmission electron microscopy (TEM) was used to confirm the crystal structure and composition of all the new phase pure and multiphase samples. The structures of single crystallites have been confirmed by selected area electron diffraction (SAED). The chemical composition has been confirmed by energy-dispersive X-ray spectroscopy (EDS). We illustrate TEM measurements for HfIrSb in Fig. 4 and for all the other materials in the Supplementary Information (Figs. S9-S23). For all nine multiphase samples (ScRhTe, ZrRhBi, ScPtBi, VRhSi, VIrSi, ScPdP, ZrIrBi, TiIrP and ZrPdPb) the diffraction pattern simulated from the predicted crystal structure enables the direct identification of the new material in a mixture of known or even unknown phases. ***The 15 previously missing never before made materials synthesized here plus the additional TaIrGe and TaCoSn synthesized earlier[38,45] all crystallized in their predicted crystal structures,*** *thus validating the theoretical procedure used*. The deviation of measured versus predicted lattice parameters (typically within 1-2%) are shown in Supplementary Information. This success rate is noteworthy given that each compound could exist in at least one of ~40 possible structure types (by analogy with other ABX cases) and that 15 out of 15 materials made adopted the predicted structure with lattice parameters extremely close to those predicted.

Some of the newly synthesized materials have compositions and structures that are nontrivial to intuitively guess. Contrary to the case of synthesis of ScPtBi that is intuitive (since its neighboring compounds ScNiBi and ScPdBi have been previously reported (Fig. 1)), the synthesis of 1:1:1 ScRhTe was not obvious since this material is the first of its kind in the III-IX-VI family in which no compound has been previously reported and in which a large majority of the compositions are predicted to be unstable (Fig. 2). Interestingly, the predicted stable ScRhTe could be synthesized but the predicted *un*stable ScIrTe could not be synthesized by similar approaches, consistent with theory. Moreover, attempts to synthesize the predicted u*nstable* VRhSn, InPdBi,

MgNiTe, NbCoPb, VCoPb, VIrSn, MgNiS and MgPdS were also unsuccessful as predicted.

***The combination of synthesis with theoretical prediction leads to acceleration of each step of the experimental discovery process:*** The first step in the discovery of new functional materials is to select the reactants and their ratios. In the past decades, chemists have developed empirical and semi-empirical rules to narrow the large number of possible Daltonian stoichiometries of elements that can be imagined. However, chemical intuition may not always hit the mark so synthetic attempts guided by such intuition may fail *without providing indications as to the source of failure.* From the prediction, useful information about the stability of the materials in comparison to their competitors can be rationalized. The instability of materials can be identified and understood while the stability of counterintuitive materials can be highlighted. It is important to note that the comparison with competitors improves the efficiency of the predictive theory and provides the experimentalist with the success rate and, in some cases, growth conditions.

Newly made solid samples are often multiphasic. Thus, most methods to synthesize new materials lead to polycrystalline samples with a mixture of known and unknown phases. Using X-Ray diffraction, the experimentalists can identify the newly made compound by isolating the diffraction peaks belonging to this phase in a deductive manner. This work is realized by identifying the other known phases of the sample and also distinguishing the peaks that cannot be indexed using the powder diffraction databases. However, the new phase cannot be identified among other unknown phases. In this context, the theoretical approach used here provides useful information: the predicted crystal structure. The diffraction pattern simulated from the predicted crystal structure enables the direct identification of the new material in a mixture of known or even unknown phases. The acceleration of identification (and also refinement) is enabled by the accuracy of the prediction concerning the symmetries (unit-cell parameters are usually less than 2-3% off).

*Functionality of new materials:* In the inverse design search for compounds with target functionality we (i) select a broad chemical group within the database of *previously made compounds*, (ii) construct a metric that reflects the target functionality, then (iii) search the database for material following this functionality metric. Left out of this procedure are materials not listed in databases because they were never made. Here we augment the database of previously made compounds by those "missing compounds" predicted to be thermodynamically stable, ensuring that one is searching for target functionality not just among the "usual suspects" materials. Furthermore, the characterization of the properties of new materials is accelerated owing to the calculation of their functionalities. The prediction will highlight the most interesting properties of the materials on which the experimentalists will focus. This information is also used in the first step of materials discovery for the selection of the predicted stable materials that will be synthesized.

*Trends in band gaps:* We have used HSE06[46] wave functions with spin-orbit coupling to evaluate the band gaps of the predicted materials. Our calculations (see Table 1) show that all 18-electron ABX compounds with non-cubic structures and cubic ABX compounds with one transition metal element are metallic. For example, the new II-X-VI (MgPdTe) and III-X-V (III = Al, Ga, In) (AlNiP, AlNiAs, AlNiSb, GaNiSb, InNiSb, and InPdSb) cubic phases with only one transition metal specie are metallic. In contrast cubic ABX compounds with two transition metals are semiconductor. This can be understood from considering the basic electronic structure of the crystal (see Fig. 5**a**):

the transition metal atoms A and B are mutually tetrahedrally coordinated nearest neighbors. Thus the *d* states of A and B, i.e. *t₂(A, d)* and *t₂(B, d)* as well as *e_g(A, d)* and *e_g(B, d)* strongly couple and repel each other, opening a large gap between the high-lying unoccupied *d* states and the low-lying occupied *d* states. There is a hybridized *a₁(s)* state residing near the band edges as illustrated in Fig. 5**a**. This *a₁(s)* state can drop down if the *s* orbitals constituting the *a₁(s)* state are low-lying. Figures 5**b-e** show the calculated band structures of TaIrSn, ZrIrSb, HfIrSb and HfIrAs illustrating the downward shifting of the *a₁(s)* state (labeled as $\Gamma_6^s$ at $\Gamma$ point) relative to the $\Gamma_8^{p,d}$ state at $\Gamma$. In HfIrAs, the $\Gamma_6^s$ state drops below the $\Gamma_8^{p,d}$ state (similarly for HfIrBi see Supplementary Fig. S3), and this band inversion leads to a topological phase transition as found for ScPtBi in another study[28] in an *assumed structure*. We have confirmed *ex post facto* the stability of that structure type and confirmed its band inversion (see Supplementary Fig. S4). The non-monotonic trends for HfIr(As, Sb, Bi) subgroup, reflecting that the drop of $\Gamma_6^s$ state can be induced by either relativistic effect of heavy elements (e.g. Hg in HgTe[47] or Bi in ScPtBi and HfIrBi) or low-lying *s* orbitals of low-Z elements (e.g. As in HfIrAs compared to Sb in HfIrSb). The predicted topological semimetals provide a platform for designing new topological insulators by application of quantum confinement (e.g. on HgTe[47]) or strain (e.g. on ScPtBi[28]).

*Thermoelectric materials:* From the 20 semiconductors predicted here (see Table 1), we find three never-before made 18-electron ABX compounds with small band gaps ($E_g < 0.5$ eV): ZrNiPb, HfNiPb and HfPdPb. Among them ZrNiPb is made single phase. The measured electron conductivity and thermopower of synthesized ZrNiPb at room temperature (RT) are 220.1 S/cm and -153.9 µV/K, respectively, giving a power factor as high as 5.2 µW/cm K² (see Supplementary Table S22), this is larger than the RT power factor of Zr₀.₅Hf₀.₅NiSn (3 µW/cm K²) with reported figure of merit[48] $ZT > 0.5$ at 700 K.

For comparison, the RT power factors of synthesized HfIrSb and ZrIrSb are 0.22 μW/cm K$^2$ and 0.013 μW/cm K$^2$, respectively. Analogous to Zr$_{0.5}$Hf$_{0.5}$NiSn, one can use alloys of ZrNiPb and HfNiPb that have very small lattice mismatch (0.2%) similar to AlAs and GaAs to reduce thermoconductivity. Indeed, total energy calculation of Zr$_{0.5}$Hf$_{0.5}$NiPb with Zr and Hf decorated on the 4 equivalent A sites in the unit cell of ABX LiAlSi-type structure shows that its formation enthalpy relative to ZrNiPb and HfNiPb is neglectable (0.4 meV/atom).

*Transparent conductors:* As a rather rare functionality among the 18-electron ABX family discovered in TaIrGe recently[38] from inverse design approach, we focus here on realizing more predicted transparent conductors. Three compounds (TiIrSb and ZrIrSb from group IV-IX-V and TaIrSn from group V-IX-IV) with wide direct gaps ($E_g^{dir}$) of 2.3~2.5 eV (similar to TaIrGe[38]) are identified as potential transparent conductors. The calculated optical absorption coefficient ($\alpha$) of TiIrSb, ZrIrSb and TaIrSn (see Supplementary Fig. S5) illustrates that the optical transition across $E_g^{dir}$ (see Table 1) is allowed and that strong optical absorption ($\alpha=10^6$ cm$^{-1}$) starts near 3 eV (for TaIrSn) or above (for TiIrSb and ZrIrSb). The strong absorption peaks above 3 eV are seen in optical absorption spectra of TiIrSb and ZrIrSb, obtained from the ultraviolet-visible diffuse reflectance measurements (see Supplementary Table S21). The measured optical gaps of TiIrSb (2.4 eV), ZrIrSb (1.9 eV) and TaIrSn (2.4 eV) are close to the predicted direct gaps as listed in Table 1. We measured the electrical conductivity of ZrIrSb as a representative of the predicted potential transparent conductors using the Van der Pauw method at room temperature. It is found that ZrIrSb is a *p*-type transparent conductor with hole conductivity as high as 6.5 S/cm (see Table S22) which is much higher than both TaIrGe (0.35 S/cm)[38] and the first *p*-type transparent conductor CuAlO$_2$ (1 S/cm)[49]. The experimental measurement on standard properties that are, however, hard to evaluate from DFT theory (e.g. conductivity) further highlights the predicted functional materials.

*Conclusions and discussion:* A systematic search of missing materials using first principles thermodynamics that is readily iterated with experimental realization (i) provides guidelines as to which of the previously overlooked materials are likely to be stable in a predicted structure with specified materials properties. It could significantly narrow the range of materials that need to be targeted experimentally for given functionalities (transparent conductors and topological semimetals). (ii) The method provides the identity of the low-lying competing phases, thereby allowing the future development of synthetic strategies that would destabilize such competing phases. (iii) The approach naturally completes the chemical series for previously missing material entries, thus establishing a sounder basis for distilling chemical rules and regularities. (iv) The first principles thermodynamics protocol we use allows us to establish in many cases the *reason* behind the instability of groups of missing compounds. (v) The approach might discourage the practice of studying theoretically exciting physical *properties* of hypothetical compounds and structures that can be shown by the current protocol to be manifestly unstable, thereby reducing experimental trial and error. (vi) The information provided by theory regarding the stable crystal structure could accelerate the laboratory identification of the structure from the measured XRD pattern. Indeed, the inverse XRD problem (going from measured XRD pattern to deduced crystal structure) is not usually straightforward for experimentalists, and often many guesses of structure type have to be attempted. Here, our theoretical result provides an excellent starting model. Fifteen missing ABX materials have been synthesized in their predicted crystal structures.

**Methods**

**Theoretical determination of crystal structure.** In this step we aim to find the structure of ABX that has the lowest total energy at $T = 0$. The total energy is calculated in the framework of density functional theory using the Vienna ab-initio simulation package (VASP)[50] (Supplementary section I where our systematic correction to DFT formation enthalpy errors and the electronic structure evaluation method are also described). The crystal structure determination is done in step (1a) by comparing the total energies of a specific ABX in a list of previously reported structure-types (Supplementary section II). The stablest structures emerging from the static stability test in step (1a) are tested in step (1b) by examination of their dynamic stability by phonon calculations using the density-functional perturbation theory (DFPT) as implemented in Quantum Espresso[51] (Supplementary section III). We find all final structures are dynamically stable, i.e. have no negative phonon frequency. The dielectric constants (including ionic contributions) are evaluated using DFPT. We have examined in a few cases that the list of structure-types used in step (1a) is not too restrictive by performing in step (1c) a genetic algorithm based structure search in which one starts from random lattice vectors and random cell positions (Supplementary section IV). Steps (1a) and (1c) are illustrated in Supplementary Fig. S1 for HfNiPb.

**Thermodynamic stability.** In this step we compare the energy of the lowest energy crystal structure of each thermodynamically stable compound with the energies of different combinations of unary, binary and ternary competing phases (described in Supplementary section V). We use the formation enthalpies ($\Delta H_f$) of all competing phases to construct a 3D "convex hull" denoted as $C(x)$ that represents the locus of the $\Delta H_f$ vs composition $x$ of the stablest structures for the system, as illustrated by the blue lines in Supplementary Fig. S1**c** for an A-X binary system for clarity. The energy (enthalpy) distance between the stablest ABX to the combination of closest competing phases as illustrated by the red arrow in Supplementary Fig. S1c is given by $\Delta H_f(ABX)-C(x_{ABX})$; this is a measure of the thermodynamic stability of ABX compound. The limitation and validation of our methodology are discussed in Supplementary sections VI and VII, respectively.

**Synthesis.** The synthesis of the ABX materials has been attempted by arc melting and silica tube annealing. Stoichiometric ABX mixtures of pure elements were pellet pressed prior to synthesis. Arc melting syntheses were performed in a Compact Arc-Melter MAM-1under argon. The annealing is realized by heating the samples in sealed silica tube under vacuum at temperatures between 500°C and 1100°C. These tubes were carbon-coated to avoid oxidation of the samples. The synthesized samples were analyzed by powder X-ray diffraction in a Rigaku Ultima IV by θ/2θ scanning with a step size of 0.02°. The selected area electron diffraction (SAED) and energy dispersive X-ray spectroscopy (EDS) experiments were conducted using a JEOL 2100 microscope operated at 200 KeV. The simulations of electron diffraction patterns were conducted using the SingleCrystal software.

**Optical measurement.** Diffuse reflectance measurements were performed using a Lambda 1050 UV/Vis/NIR spectrophotometer with an integrating sphere attachment (PerkinElmer, Oak Brook, IL). Spectra were taken from 250 to 2500 nm and two baseline spectra, at 0% and 100% R, were taken using pressed polytetrafluoroethylene (PTFE) powder compacts. The diffuse reflectance spectra were converted to optical absorption spectra by the Kubelka-Munk relation[52] and the band gaps estimated by Tauc plots[53].

**Electrical measurement.** The conductivity and thermopower of sintered materials were measured at ambient conditions by methods described in detail by Hong, et al.[54] The measured conductivities were corrected for porosity by the Bruggemann symmetric model[55] and for the geometry of the sample as outlined by Smits, et al.[56]

**Acknowledgements**

This work was performed at Northwestern University and Colorado University with funding provided by the U.S. Department of Energy, Office of Science, Basic Energy Sciences, and Energy Frontier Research Centers, under Contract No. DE- AC36-08GO28308 to NREL. We thank Giancarlo Trimarchi for helpful discussions.

**Author Contributions**

XZ performed the stability calculations as well as property calculations other than the phonon and dielectric constants done by LY. RG, YL, LH, TS and DC performed the synthesis and characterization of new ABX materials. The experimental work was supervised by KRP. Furthermore, XZ, RG and LY contributed to the writing of the paper. KRP and AZ directed the analysis of the results and writing of the paper.

**Competing financial interests**

The authors declare no competing financial interests.

**Figures**

**Figure 1 | ABX compounds in N-X-(8-N) groups (N=I, II, III, IV).** Check marks: previously reported compounds[1-3,34-36]; plus: unreported and predicted here to be stable; minus: unreported and predicted here to be unstable; circle: too close to call. The compounds that have the cubic half-Heusler structure (LiAlSi-type, F-43m) are denoted by the bold violet symbols. The existing compounds appear in 7 structure types; a total of 41 structure types were examined theoretically.

**Figure 2 | ABX compounds in (N+1)-IX-(8-N) (N=I, II, III, IV) groups.** Check marks: previously reported[1-3,34-36]; plus: unreported and predicted here to be stable; minus: unreported and predicted here to be unstable; circle: too close to call. The compounds that have the cubic half-Heusler structure (LiAlSi-type, F-43m) are denoted by the bold violet symbols.

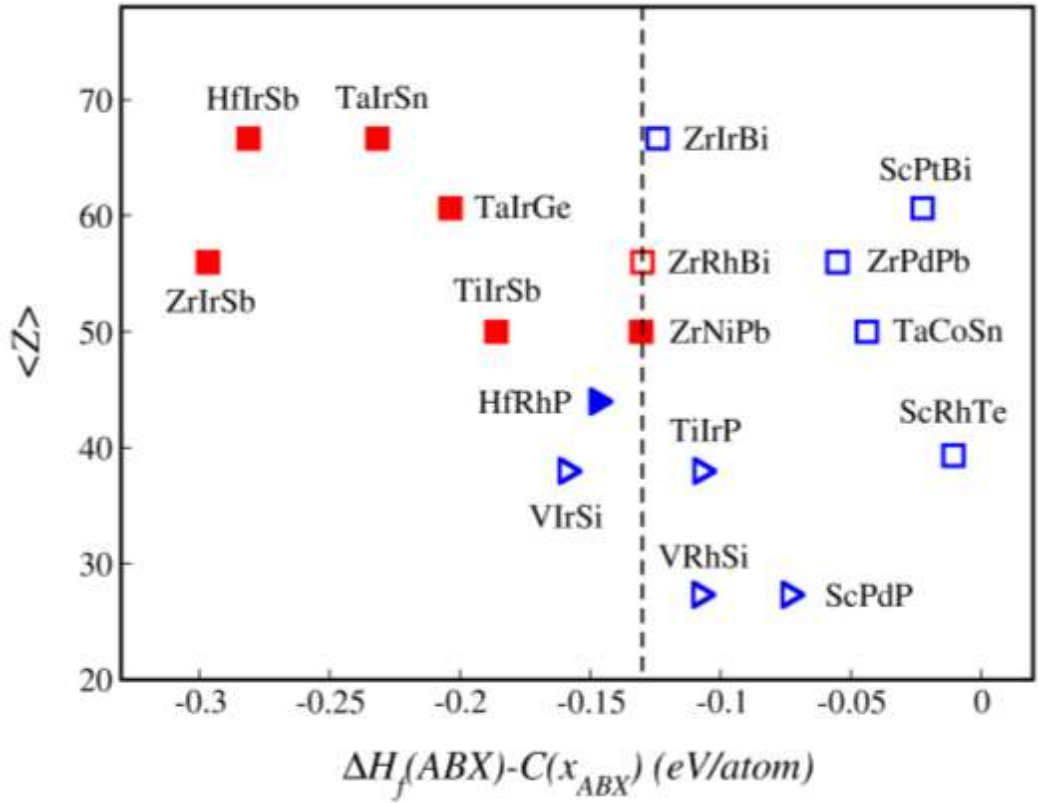

**Figure 3 | 18-electron ABX compounds** predicted by theory and synthesized by Experiment in this work. $\Delta H_f(ABX)-C(x_{ABX})$ is the energy distance to the strongest competing phases (or convex hull see Methods section) and $<Z>$ is the average atomic number. Square: in cubic LiAlSi-type (F-43m) structure. Triangle: in orthorhombic MgSrSi-type (Pnma) structure. Solid symbol: single phase. Empty symbol: multi phase. VIrSi and ZrRhBi are multi phases with high purity (see Supplementary Information). Dashed line (at $\Delta H_f(ABX)-C(x_{ABX}) = -0.13$ eV/atom): diagrammatical separation of single phase and multi phase compounds (one error out of 15 cases). Red symbol: with measured properties (see Table S21 and S22). Blue symbol: properties not measured.

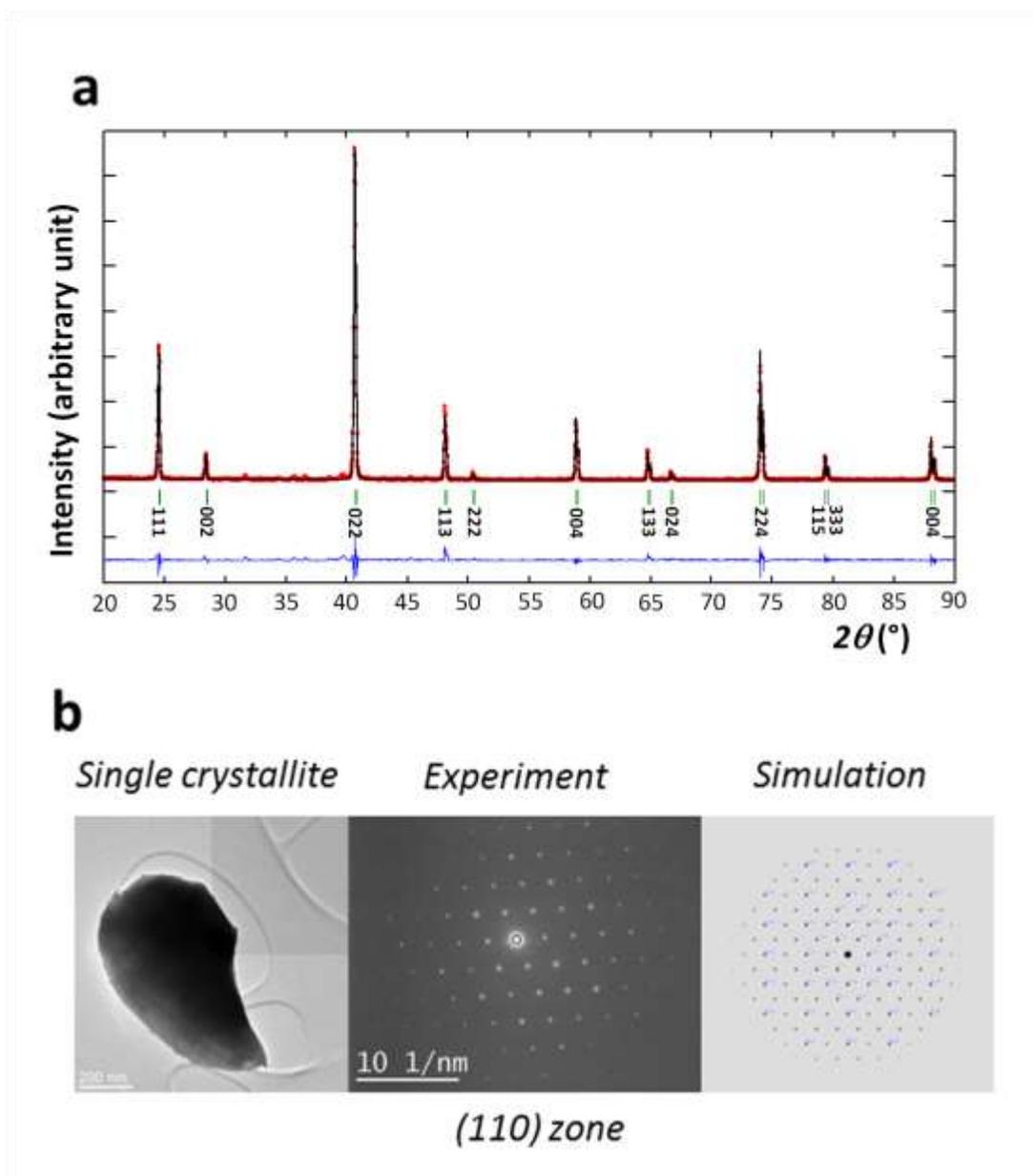

**Figure 4 | Discovery of HfIrSb ternary material. a,** Rietveld refinement carried out on the pure sample obtained after annealing under vacuum at 1050°C. Red point: observed X-ray diffraction intensity; black curve: calculated intensity; blue curve: difference between observed and calculated intensities; green line: Bragg position. **b,** Selected area electron diffraction (SAED) and energy dispersive X-ray spectroscopy (EDS) further confirm the predicted crystal structure (LiAlSi-type, F-43m) (see more details in Supplementary Information).

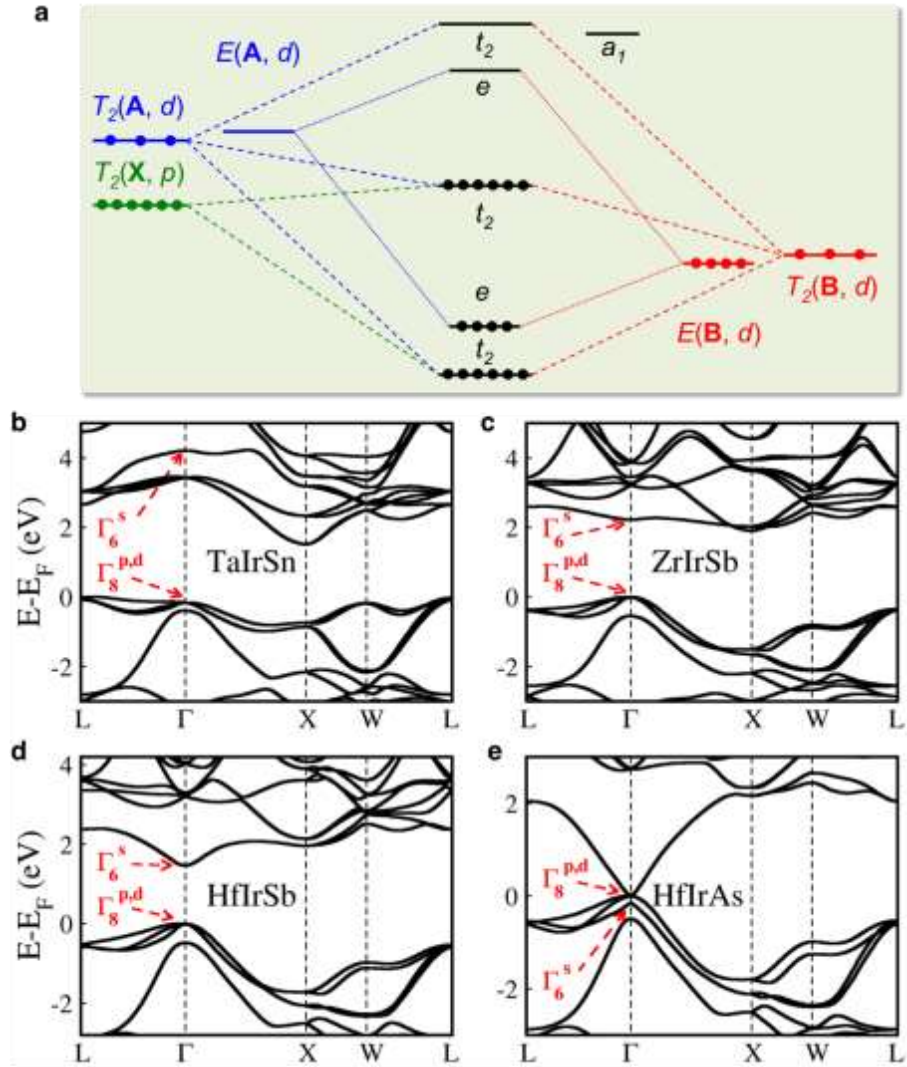

**Figure 5 | Electronic structure of 18-electron ABX in the cubic half-Heusler structure (LiAlSi-type, F-43m). a,** Schematic energy level diagram of 18-electron ABX half-Heuslers. In the cubic structure (LiAlSi-type), all atomic sites have $T_d$ symmetry, thus the $d$ orbitals split into $e$ and $t_2$ states, and the latter couples with the $t_2$ states of $p$ orbitals. The $s$ orbitals of all atoms are hybridized in the $a_1$ states. The high lying $p$ orbitals of A and B atoms are not shown. **b-e,** Band structures of TaIrSn, ZrIrSb, HfIrSb and HfIrAs from HSE06 with SOC.

**Table 1:** Lowest-energy structure, lattice constant (in Å), formation enthalpy ($\Delta H_f$ in eV/atom) from DFT, fundamental band gap ($E_g$ in eV), direct gap ($E_g^{dir}$) and spin splitting of the first valence and conduction bands at $W$ point ($\Delta_{VB1}^W$ and $\Delta_{CB1}^W$ in eV) from HSE06 with SOC, as well as dielectric constant ($\varepsilon_0$) from DFPT of the predicted stable ABX semiconductors or semimetals. The predicted stable metals are distributed in 8 structure types (see Tables S1, S2, and S19): LiAlSi-type (MgPdTe, AlNiP, AlNiAs, AlNiSb, GaNiSb, InNiSb, and InPdSb), MgSrSi-type (GaPtSb, InPtSb, ScNiAs, ScPdP, ScPdAs, YNiAs, YPdAs, LaPtAs, TiRhP, TiIrP, HfRhP, HfRhAs, VRhSi, VRhGe, VIrSi, and VIrGe), ZrBeSi-type (YPdP and LaNiAs), CaPdSi-type (GaNiBi and AlIrSe), CoYC-type (LaNiN), LiBeN-type (GaIrTe), AuEuGe-type (HfIrP), and SmSI-type (AlPtSb).

| Compounds | Structure (space group) | Lattice constant | $\Delta H_f$ | $E_g$ ($E_g^{dir}$) | $\varepsilon_0$ |
|---|---|---|---|---|---|
| ScPtBi | s1[a] (F-43m) | 6.557 | -0.86 | 0.00 (0.00) | 157.90 |
| TiPdSn | s1 (F-43m) | 6.230 | -0.52 | 0.74 (1.59) | 21.61 |
| ZrNiPb | s1 (F-43m) | 6.267 | -0.65 | 0.43 (0.87) | 22.22 |
| ZrPdPb | s1 (F-43m) | 6.506 | -0.68 | 0.53 (1.09) | 21.67 |
| ZrPtPb | s1 (F-43m) | 6.518 | -0.84 | 1.01 (1.93) | 19.74 |
| HfNiPb | s1 (F-43m) | 6.252 | -0.57 | 0.32 (1.05) | 22.32 |
| HfPdPb | s1 (F-43m) | 6.485 | -0.59 | 0.49 (1.30) | 22.51 |
| HfPtPb | s1 (F-43m) | 6.485 | -0.77 | 0.97 (1.31) | 21.41 |
| ScRhTe | s1 (F-43m) | 6.350 | -1.05 | 0.75 (0.75) | 17.87 |
| TiIrAs | s1 (F-43m) | 5.941 | -0.70 | 1.40 (1.40) | 17.54 |
| TiIrSb | s1 (F-43m) | 6.169 | -0.76 | 1.63 (2.39) | 16.05 |
| ZrRhBi | s1 (F-43m) | 6.462 | -0.71 | 1.28 (1.59) | 17.93 |
| ZrIrAs | s1 (F-43m) | 6.182 | -0.99 | 0.78 (0.78) | 18.41 |
| ZrIrSb | s1 (F-43m) | 6.372 | -1.11 | 1.91 (2.25) | 15.11 |
| ZrIrBi | s1 (F-43m) | 6.496 | -0.76 | 0.71 (0.71) | 20.83 |
| HfRhBi | s1 (F-43m) | 6.455 | -0.64 | 0.60 (0.60) | 23.29 |
| HfIrAs | s1 (F-43m) | 6.159 | -0.96 | 0.00 (0.00) | 115.13 |
| HfIrSb | s1 (F-43m) | 6.345 | -1.07 | 1.49 (1.49) | 15.45 |
| HfIrBi | s1 (F-43m) | 6.476 | -0.71 | 0.00 (0.00) | 89.32 |
| TaCoSn | s1 (F-43m) | 5.974 | -0.29 | 1.37 (1.63) | 20.13 |
| TaRhSn | s1 (F-43m) | 6.201 | -0.48 | 1.40 (1.76) | 18.32 |
| TaIrGe | s1 (F-43m) | 6.026 | -0.75 | 1.62 (2.49) | 16.86 |
| TaIrSn | s1 (F-43m) | 6.233 | -0.67 | 1.55 (2.26) | 16.93 |

[a]LiAlSi-type structure.